# Imaging the Meissner effect in pressurized bilayer nickelate with integrated multi-parameter quantum sensor

Junyan Wen[1,2†], Yue Xu[1,2†], Gang Wang[1,2†], Ze-Xu He[1,2], Yang Chen[1], Ningning Wang[1,2], Tenglong Lu[1], Xiaoli Ma[1,2], Feng Jin[1,2], Liucheng Chen[1], Miao Liu[1], Jing-Wei Fan[3], Xiaobing Liu[4], Xin-Yu Pan[1,2,5], Gang-Qin Liu[1,2,5*], Jinguang Cheng[1,2*], Xiaohui Yu[1,2*]

[1]*Beijing National Laboratory for Condensed Matter Physics and Institute of Physics, Chinese Academy of Sciences, Beijing 100190, China*
[2]*School of Physical Sciences, University of Chinese Academy of Sciences, Beijing 100190, China*
[3] *Department of Physics, Hefei University of Technology, Hefei, Anhui 230009, China*
[4] *Laboratory of High Pressure Physics and Material Science, School of Physics and Physical Engineering, Qufu Normal University, Qufu, Shandong 273165, China*
[5] *CAS Center of Excellence in Topological Quantum Computation, Beijing 100190, China*
† These authors contribute equally to this work.
*Corresponding authors: gqliu@iphy.ac.cn; jgcheng@iphy.ac.cn; yuxh@iphy.ac.cn

## ABSTRACT

Recent reports on the signatures of high-temperature superconductivity with a critical temperature $T_c$ close to 80 K have triggered great research interest and extensive follow-up studies. Although the zero resistance has been successfully achieved under improved hydrostatic pressure conditions, the Meissner effect of $La_3Ni_2O_{7-\delta}$ under high pressure remains controversial. Here, using shallow nitrogen-vacancy centers implanted on the culet of diamond anvils as in-situ quantum sensors, we observe compelling evidence for the Meissner effect in polycrystalline bilayer nickelate samples: the magnetic field expulsion during both field cooling and field warming processes. In particular, we explore the multiparameter measurement capacity of the diamond quantum sensors to extract the weak demagnetization signal of $La_3Ni_2O_{7-\delta}$. The correlated measurements of Raman spectra and magnetic imaging indicate an incomplete structural transformation related to the displacement of oxygen ions emerging in the non-superconducting region. Our work clarifies the controversy about the Meissner effect of $La_3Ni_2O_{7-\delta}$ and contributes to the development of quantum sensing of weak signals under high-pressure conditions.

# INTRODUCTION

Since superconductivity has been observed in thin $Nd_{1-x}Sr_xNiO_2$ films, the study of nickelate superconductors has attracted substantial attention in condensed matter physics community [1-5]. Notably, pressure plays a significant role in regulating the superconductivity of nickelates. It has been shown that the $T_c$ of $Pr_{0.82}Sr_{0.18}NiO_2$ thin films can be significantly increased to ~ 31 K by applying hydrostatic pressure up to 12 GPa [6]. Recent studies have observed signatures of high-temperature superconductivity (HTSC) in pressurized single crystals of $La_3Ni_2O_{7-\delta}$ with an onset $T_c$ of about 78 K, exceeding the McMillan limit and indicating a potential new class of HTSC at liquid nitrogen temperatures [7]. This significant discovery rapidly garnered widespread attention within the superconductivity research community [8-12]. With improved hydrostaticity, zero resistance has been researched [13-15], and ARPES [16,17] and optical conductivity [10] have been used to characterize its electronic properties. Meanwhile, inelastic X-ray scattering (RIXS) [18], muon-spin relaxation ($\mu$SR) [19,20] and nuclear magnetic resonance (NMR) [21] have been used to reveal the pressure-driven transition of SDW and CDW in nickelates.

In contrast to the extensive and unambiguous evidence of its electronic properties, the final criterion of superconductivity, the Meissner effect, remains controversial for $La_3Ni_2O_{7-\delta}$ under pressure [7,22]. Using modulated a.c. susceptibility measurements, Zhou *et al* found that the relative superconducting volume fraction in $La_3Ni_2O_7$ is only 0.68%, suggesting that the superconductivity in this nickelate is filamentary-like [23]. Using SQUID magnetometry, Li *et al* reported that the maximum superconducting volume fraction of $La_3Ni_2O_7$ at 22.0 GPa reaches 62.7%, suggesting the bulk nature of superconductivity [24]. Such a large difference in superconducting volume fraction may result from a mixture of $La_3Ni_2O_7$ and other phases such as $La_4Ni_3O_{10}$ and $La_2NiO_4$. Indeed, using multislice electron ptychography and electron energy-loss spectroscopy, Li *et al* found considerable inhomogeneity of oxygen content within the sample [25]. By replacing one-third of the La with Pr, Wang *et al* demonstrated that the trilayer phase can be significantly suppressed, resulting in a superconducting volume fraction of more than 57% in $La_2PrNi_2O_7$ [26]. Accordingly, it is important to re-examine the Meissner effect of the bilayer nickelate $La_3Ni_2O_7$.

In this work, we adapt the newly developed quantum sensing techniques with nitrogen-vacancy (NV) centers integrated into the diamond anvil cells (DAC) [27-32], assisted by correlated Raman spectra, to image the Meissner effect of bilayer nickelate under high pressure. With the spatially resolved magnetic field imaging around the nickelate samples, clear magnetic field expulsion effects are observed on a scale of tens of micrometers, indicating the bulk nature of superconductivity. In particular, to extract the weak demagnetization signal from the pressure-sensitive $La_3Ni_2O_7$ sample, we explore the multiparameter measurement capacity of the NV center and demonstrate the simultaneous imaging of pressure and magnetic field under high pressures. The correlated measurements of Raman spectra and NV-based magnetic imaging indicate an incomplete structural transformation related to the displacement of oxygen ions emerging in the non-superconducting region. Our results provide crucial evidence for the Meissner effect in pressurized bilayer nickelate superconductors and contribute both to the mechanistic understanding of their HTSC and to the development of NV-based quantum sensing under high-pressure conditions.

## Results

As shown in Fig. 1a, the high-pressure sample chamber consists of two diamond anvils and a gasket. Through the transparent diamond window, optically detected magnetic resonance (ODMR) and Raman spectra can be measured under high pressure, in a correlated manner. To investigate the magnetic field and stress distribution around the superconducting samples, shallow NV centers are created on one of the diamond culets by nitrogen ion implantation (20 keV, $2 \times 10^{14}$ cm$^{-2}$) and subsequent high temperature annealing (800 °C, 2 hours). The spin resonance frequencies of an NV center are sensitive to the local magnetic field (and also to pressure, temperature and so on) and can be read out optically, providing a convenient and efficient method to study magnetic properties inside DACs. Details of the ODMR technique can be found in the methods and previous works [29,33].

The Meissner effect of $La_2PrNi_2O_7$ and $La_3Ni_2O_{7-\delta}$ polycrystalline samples was studied using diamond quantum sensors, in conjunction with in-situ Raman measurements. For sample A,

La$_2$PrNi$_2$O$_7$ in liquid pressure transmitting medium (silicon oil), ODMR spectra were measured under different pressures and external magnetic fields. We also performed experiments with La$_2$PrNi$_2$O$_7$ (sample B) in a solid pressure transmitting medium (KBr) to investigate the influence of hydrostatic pressure conditions on superconductivity. Correlated Raman and ODMR measurements were performed on sample B to determine the structural distinctions on both the superconducting and non-superconducting regions. Finally, sample C, La$_3$Ni$_2$O$_{7-\delta}$ in silicon oil, was measured. Since the diamagnetism signal of this La$_3$Ni$_2$O$_{7-\delta}$ sample is weak and buried in the pressure gradient background, we performed additional zero-field ODMR measurements to extract and decouple the contribution of local stress, and successfully obtained the diamagnetism image of the La$_3$Ni$_2$O$_{7-\delta}$ sample.

**Local diamagnetism in La$_2$PrNi$_2$O$_7$**

We start with the sample of La$_2$PrNi$_2$O$_7$ in silicon oil (sample A) at 20 GPa. As shown in Fig. 1b and Fig. 2a, the sample edge and its relative position on the diamond culets were determined by comparing the brightfield image and the confocal NV fluorescence image, see Fig. S1 for more details. Fig. 2b shows typical ODMR spectra of NV centers near the sample. These spectra were measured under an external magnetic field of around 120 G after zero-field cooling the sample to 6 K. The strength of the external magnetic field was calibrated using the ODMR splitting of the NV centers away from the sample (point A0 in Fig. 2a). For NV centers directly above the sample, e.g. at point A1, the ODMR splitting (644.1 MHz) is noticeably smaller than that of the reference point (672.9 MHz), indicating local diamagnetism of the La$_2$PrNi$_2$O$_7$ sample. At the same time, a relatively larger splitting (678.0 MHz) was observed at point A2, which was due to the magnetic flux concentration at the sample edge. By measuring more points around the sample, a superconducting region can be identified, as shown by the blue color in Fig. 2c. For the measured region of 110 μm * 74.5 μm, about 2/5 of the points show diamagnetism, indicating a relatively large amount of superconducting shielding volume in the La$_2$PrNi$_2$O$_7$ sample. In addition, Figure 2d presents diamagnetism (at point A1) and flux concentration effects (at point A2) under different external magnetic fields, which show an almost linear dependence in all measured magnetic fields.

To further verify the Meissner effect of the La$_2$PrNi$_2$O$_7$ sample, more ODMR spectra were acquired during zero-field-cooling-field-warming (ZFC-FW) and field-cooling (FC) processes, with an external magnetic field of 120 G. The detailed experimental protocol is shown in Fig. 2e.

As displayed in Fig. 2f, both the diamagnetism (at point A1) and the flux concentration (at point A2) show a clear temperature dependence. In comparison, the reference point (A0) remains almost constant splitting throughout the ZFC-FW and FC processes. From these results, a superconducting transition temperature $T_c \sim 60$ K is obtained and the expulsion of magnetic flux throughout the FC process, an intrinsic property of the Meissner effect, is observed. Meanwhile, compared to the ZFC-FW process, a weaker diamagnetism effect was observed at point A1 during the FC process, which can be attributed to the flux trapping effect [31].

To rule out the possibility that the diamagnetism originates from other magnetic impurities, both the ZFC-FW and FC measurements were performed under an external magnetic field of opposite direction and a strength of about -210 G (see Fig. 2e for the experimental protocol). The diamagnetism effect and its temperature dependence are similar to those observed at 120 G (shown in Fig. S2), indicating that the local diamagnetism sensed by the NV centers is not due to the compensating effect of local magnetic impurities in or near the samples. In addition, we varied the pressure from 11 GPa to 30 GPa and performed ZFC-FW measurements. The reduced diamagnetic strength at pressures above 20 GPa indicates that superconductivity is gradually suppressed by the pressure (shown in Fig. S3). The $T_c$ determined from the ODMR measurements are included in the phase diagram in Fig. S3d. After the pressure decreased to 11 GPa, the effect of diamagnetism disappeared, which is consistent with the electrical resistance results in the previous study [26].

**Simultaneous magnetic and Raman measurements**

To investigate the effects of pressure transmitting media, another $La_2PrNi_2O_7$ sample (sample B) was loaded in KBr (a solid pressure transmitting medium) and compressed to 30 GPa. We then characterized the sample with both ODMR and Raman meassurements. As shown in Fig. 3a, the confocal image clearly shows the shape and relative position of the sample, the microwave wire, and the edge of the sample chamber. Fig. 3b shows a magnetic field image under an external magnetic field of about 34 G after zero-field cooling of the sample to 7 K. The test area of this sample is limited to the area next to the microwave antenna to maintain a good ODMR contrast at low microwave power. Local diamagnetism (reduction of ODMR splitting) was observed in two regions of the sample. The local diamagnetism of one region was tracked during the FW process (under an external magnetic field of around 120 G), as shown in Fig. 3c. As the temperature increases to $T_c$, the diamagnetic effect gradually fades out. When the temperature is above $T_c$, the entire region exhibits a uniform magnetic field in line with the strength at reference point B0. More results of the ZFC-FW measurement in sample B can be found in Fig. S4. Compared to sample A, the diamagnetism effect in sample B is smaller. This is plausible because a liquid pressure transmitting medium (silicon oil) was used in sample A therefore a better hydrostatic condition was obtained. As a result, a larger superconducting volume and more pronounced diamagnetism was observed in sample A. These results are consistent with the previous studies on the electrical resistance [7,13,14,34].

Thanks to the high spatial resolution of the NV-based magnetometer, the distribution of the superconducting regions is visually obtained, which enables us to perform targeted spectroscopic measurements. We then used in-situ Raman spectrum to study the structure of pressurized $La_2PrNi_2O_7$ in both the superconducting and non-superconducting regions, which are marked in Fig. 3b. As shown in Fig. 3d and Fig. S5b, a slight difference is observed in the Raman spectra, where a satellite peak around 680 cm$^{-1}$ is suppressed in the superconducting region at 20 GPa. The satellite peak is more pronounced in the non-superconducting region and in the sample before compression (see Fig. S6a).

Our density-functional perturbation theory (DFPT) calculations show that structural transformations in $La_3Ni_2O_7$ manifest as degeneracy of Raman peaks. When $La_3Ni_2O_7$ is compressed from 0 GPa to about 9 GPa, the difference between the $B_{2g}$ and $B_{1g}^2$ modes gradually becomes negligible, and a single mode ($E_g$) appears under high pressures, as shown in Fig. S6b. The same phenomenon is observed for $Pr_3Ni_2O_7$ (Fig. S6c), implying that the above discussion may reflect the common feature of this class of materials. Indeed, a similar phenomenon has been observed for the bilayer R-P perovskite $Li_2CaTa_2O_7$, which has a similar structure to $La_2PrNi_2O_7$ [35]. A comparison with the Raman spectra of related compounds [36] shows that the satellite peak in $La_2PrNi_2O_7$ can be assigned to an oxygen $B_{1g}$ mode. Therefore, the satellite peak in the non-superconducting region may be attributed to the incomplete structural transformation caused by the displacements of the oxygen ions, which is challenging to be observed in X-ray and neutron diffraction [37]. These results are well consistent with the change of the bond angle of the Ni-O-Ni in $La_3Ni_2O_7$. In the case of $La_3Ni_2O_7$, the emergence of HTSC is along with the structural transformations from *Amam* to *Fmmm* with the change of the bond angle of the Ni-O-Ni from 168.0° to 180° along the c axis [34]. As a consequence, the electronic interactions between the bilayer of $NiO_2$ are increased, corresponding to the metallization of the inter-layer σ-bonding bands. This phenomenon is observed in the conventional high-$T_c$ superconductors, such as $MgB_2$ and $Li_3B_4C_2$ [7].

**Local diamagnetism imaging in $La_3Ni_2O_{7-\delta}$**

To investigate the Messiner effect of $La_3Ni_2O_{7-\delta}$, a polycrystalline sample was loaded with silicon oil as the pressure transmitting medium (sample C). Fig. 4a shows the confocal image of NV fluorescence when the sample is compressed to 20 GPa. Following the experimental protocols shown in Fig. 2e, the ZFC-FW and FC measurements were performed under an external magnetic field of around 120 G. As shown in Fig. 4b, local diamagnetism was observed during both the FW and FC processes. However, the diamagnetism signal (change of ODMR splitting below and above $T_C$) in sample C is only one-fifth of sample A, indicating a relatively small superconducting volume in this sample.

Notably, there are obvious differences of ODMR splitting among different NV centers, even when the temperature is above $T_c$. This phenomenon can be explained by the inhomogeneous distribution of stress on the diamond culet. In particular, the compressive stress can shift the center frequency of the ODMR spectra, while the differential stress regulates the ODMR splitting. It is worth noting that this phenomenon also occurs in the former two samples, but the contribution of stress is much weaker compared to the contribution of the superconducting diamagnetism effect in these two samples. Fortunately, diamond NV centers can also serve as in-situ stress sensors. To decouple the contribution of stress, we perform ODMR measurements at 150 K (above $T_c$) in zero field (see Supplementary Information for details). As shown in Fig. 4c and 4d, the maximum pressure variance among the sample is about 3 GPa. With this in hand, we can eliminate the influence of non-uniform stress distribution and obtain a strain-free magnetism distribution, as shown in Fig. 4e. In short, the Meissner effect of $La_3Ni_2O_{7-\delta}$ was imaged with NV centers under high pressure.

Finally, the diamagnetism signals of three samples during ZFC-FW are normalized and compared with each other. As shown in Fig. 4f, it can be clearly seen that pressurized $La_2PrNi_2O_7$ in silicon oil exhibits the strongest diamagnetism effect, while $La_2PrNi_2O_7$ in KBr has a relatively weaker signal, indicating that the hydrostatic conditions play an important role in achieving a high superconducting volume fraction. Overall, $La_3Ni_2O_{7-\delta}$ in silicon oil has the weakest signal, which is consistent with its inhomogeneous nature (mixture of $La_3Ni_2O_7$ and other phases such as $La_4Ni_3O_{10}$ and $La_2NiO_4$) [24,25].

## DISCUSSION

It is noteworthy that the superconducting diamagnetism observed in our work exhibits a weaker magnitude compared to other superconducting systems (e.g., cuprate and hydride superconductors). We attribute this weaker diamagnetic response primarily to the limited superconducting volume fraction in the measured nickelate samples. The same situation has occurred in previous electrical and magnetic measurements of nickelate, which have consistently reported relatively weak superconducting signals with pronounced sample inhomogeneity [7,11,13,14,26]. Although there is no perfect diamagnetic signal, the characteristic field-dependent magnetic response and the expulsion of magnetic flux during field-cooling measurements clearly confirm the presence of the Meissner effect in pressurized bilayer nickelate.

In conclusion, the Meissner effect is imaged in pressurized bilayer nickelate superconductors $La_3Ni_2O_{7-\delta}$ and $La_2PrNi_2O_7$ using NV center quantum sensors. By combining ZFC-FW measurements with FC measurements in different external magnetic fields, our results confirm the existence of bulk HTSC in $La_3Ni_2O_{7-\delta}$ and $La_2PrNi_2O_7$. Through the correlated measurement between the ODMR and Raman spectroscopy, we identify the structural difference between superconducting and non-superconducting regions in the $La_2PrNi_2O_7$ sample. These measurements indicate that the inhomogeneous superconductivity may be due to the incomplete structural transformations related to the displacement of oxygen ions on the non-superconducting regions and the complete transformations on the superconducting regions, which helps us to understand the underlying mechanism in nickelate high-temperature superconductors.

Our research demonstrates that diamond NV center is a powerful probe to study inhomogeneous samples under high pressure, due to their high sensitivity, high spatial resolution, and multi-parameter sensing capacity. When measuring samples with weak magnetic signals under high pressure, the influence of the stress gradient is a critical issue that should be given sufficient attention. With spatially resolved ODMR measurement, various in-situ high-pressure experiments, such as Raman and absorption spectroscopy, can be performed efficiently. Our work not only demonstrates the feasibility of using NV center sensors to measure weak magnetic signals under high pressure but also provides a detailed experimental methodology.

## SUPPLEMENTARY DATA

Supplementary data are available at NSR online.

## ACKNOWLEDGEMENTS

We thank the high-pressure synergetic measurement station of Synergtic Extreme Condition User Facility for the help in high-pressure Raman measurements.

## FUNDING

This work was supported by the National Key R&D Program of China (2023YFA1608900, 2024YFA1611300, 2021YFA1400300, 2019YFA0308100, 2023YFA1406100), the National Natural Science Foundation of China (12375304, 12404163, 12022509, T2121001, 12025408, 12404179, 12074422, 12374468), the Beijing Natural Science Foundation (Z200009, Z230005), the Innovation Program for Quantum Science and Technology (2023ZD0300600), the Strategic Priority Research Program of CAS (XDB33000000).

## AUTHOR CONTRIBUTIONS

G.Q.L., J.G.C. and X.H.Y. designed and supervised this project. G.W., N.N.W. and J.G.C. synthesized and provided the samples; J.W.F. and X.B.L. implanted NV centers into the diamond culets; J.Y.W. load the samples into high pressure chambers; Y.X., Z.X.H. and C.Y. conduct the ODMR measurements; J.Y.W., X.L.M., F.J. and L.C.C. measured the in-situ high pressure and low temperature Raman spectra; J.Y.W., Y.X., Z.X.H., C.Y., G.Q.L and X.H.Y. analyzed all the collected data; T.L.L. and M.L. performed DFT calculations and gave advice from a theoretical perspective; J.Y.W., G.Q.L., J.G.C., X.H.Y. wrote the paper with inputs from all coauthors.

*Conflict of interest statement.* None declared.

# REFERENCES

1. Li D, Lee K, Wang BY *et al.* Superconductivity in an infinite-layer nickelate. *Nature* 2019;**572**(7771):624-7.

2. Zeng S, Tang CS, Yin X *et al.* Phase diagram and superconducting dome of infinite-layer $Nd_{1-x}Sr_xNiO_2$ thin films. *Phys Rev Lett* 2020;**125**(14):147003.

3. Zeng S, Li C, Chow LE *et al.* Superconductivity in infinite-layer nickelate $La_{1-x}Ca_xNiO_2$ thin films. *Sci Adv* 2022;**8**(7):eabl9927.

4. Ding X, Tam CC, Sui X *et al.* Critical role of hydrogen for superconductivity in nickelates. *Nature* 2023;**615**(7950):50-5.

5. Osada M, Wang BY, Lee K *et al.* Phase diagram of infinite layer praseodymium nickelate $Pr_{1-x}Sr_xNiO_2$ thin films. *Phys Rev Mater* 2020;**4**(12):121801.

6. Wang NN, Yang MW, Yang Z *et al.* Pressure-induced monotonic enhancement of $T_c$ to over 30 K in superconducting $Pr_{0.82}Sr_{0.18}NiO_2$ thin films. *Nat Commun* 2022;**13**(1):4367.

7. Sun H, Huo M, Hu X *et al.* Signatures of superconductivity near 80 K in a nickelate under high pressure. *Nature* 2023;**621**(7979), 493-8.

8. Christiansson V, Petocchi F, Werner P *et al.* Correlated electronic structure of $La_3Ni_2O_7$ under pressure. *Phys Rev Lett* 2023;**131**(20), 206501.

9. Zhang Y, Lin L-F, Moreo A *et al.* Structural phase transition, s±-wave pairing, and magnetic stripe order in bilayered superconductor $La_3Ni_2O_7$ under pressure. *Nat Commun* 2024;**15**(1):2470.

10. Liu Z, Huo M, Li J *et al.* Electronic correlations and partial gap in the bilayer nickelate $La_3Ni_2O_7$. *Nat Commun* 2024;**15**(1):7570.

11. Zhu Y, Peng D, Zhang E *et al.* Superconductivity in pressurized trilayer $La_4Ni_3O_{10-\delta}$ single crystals. *Nature* 2024;**631**(8021):531-6.

12. Wang L, Li Y, Xie S-Y *et al.* Structure responsible for the superconducting state in $La_3Ni_2O_7$ at high-pressure and low-temperature conditions. *J Am Chem Soc* 2024;**146**(11):7506-14.

13. Hou J, Yang P-T, Liu Z-Y *et al.* Emergence of High-Temperature Superconducting Phase in Pressurized $La_3Ni_2O_7$ Crystals. *Chin Phys Lett* 2023;**40**(11):117302.

14. Zhang Y, Su D, Huang Y *et al.* High-temperature superconductivity with zero resistance and strange-metal behaviour in $La_3Ni_2O_{7-\delta}$. *Nat Phys* 2024;**20**:1269-73.

15. Wang G, Wang NN, Shen XL *et al.* Pressure-Induced Superconductivity In Polycrystalline $La_3Ni_2O_{7-\delta}$. *Phys Rev X* 2024;**14**(1):011040.

16. Yang J, Sun H, Hu X *et al.* Orbital-dependent electron correlation in double-layer nickelate $La_3Ni_2O_7$. *Nat Commun* 2024;**15**(1):4373.

17. Abadi S, Xu K-J, Lomeli EG *et al.* Electronic Structure of the Alternating Monolayer-Trilayer Phase of $La_3Ni_2O_7$. *Phys Rev Lett* 2025;**134**(12):126001.

18. Xie T, Huo M, Ni X *et al.* Strong interlayer magnetic exchange coupling in $La_3Ni_2O_{7-\delta}$ revealed by inelastic neutron scattering. *Sci Bull* 2024;**69**(20):3221-7.

19. Khasanov R, Hicken TJ, Gawryluk DJ *et al.* Pressure-enhanced splitting of density wave transitions in $La_3Ni_2O_{7-\delta}$. *Nat Phys* 2025;**21**(3):430-6.

20. Chen K, Liu X, Jiao J *et al.* Evidence of Spin Density Waves in $La_3Ni_2O_{7-\delta}$. *Phys Rev Lett* 2024;**132**(25):256503.

21  Zhao D, Zhou Y, Huo M *et al.* Pressure-enhanced spin-density-wave transition in double-layer nickelate La$_3$Ni$_2$O$_{7-\delta}$. *Sci Bull* 2025;**70**(8):1239-45.

22  Puphal P, Reiss P, Enderlein N *et al.* Unconventional Crystal Structure of the High-Pressure Superconductor La$_3$Ni$_2$O$_7$. *Phys Rev Lett* 2024;**133**(14):146002.

23  Zhou Y, Guo J, Cai S *et al.* Investigations of key issues on the reproducibility of high-$T$c superconductivity emerging from compressed La$_3$Ni$_2$O$_7$. *Matter Radiat Extremes* 2025;**10**(2):027801.

24  Li J, Peng D, Ma P *et al.* Identification of the superconductivity in bilayer nickelate La$_3$Ni$_2$O$_7$ upon 100 GPa. *Natl Sci Rev* 2025;nwaf220.

25  Dong Z, Huo M, Li J *et al.* Visualization of oxygen vacancies and self-doped ligand holes in La$_3$Ni$_2$O$_{7-\delta}$. *Nature* 2024;**630**(8018):847-52.

26  Wang N, Wang G, Shen X *et al.* Bulk high-temperature superconductivity in pressurized tetragonal La$_2$PrNi$_2$O$_7$. *Nature* 2024;**634**(8034):579-84.

27  Hsieh S, Bhattacharyya P, Zu C *et al.* Imaging stress and magnetism at high pressures using a nanoscale quantum sensor. *Science* 2019;**366**(6471):1349-54.

28  Lesik M, Plisson T, Toraille L *et al.* Magnetic measurements on micrometer-sized samples under high pressure using designed NV centers. *Science* 2019;**366**(6471):1359-62.

29  Shang Y-X, Hong F, Dai J-H *et al.* Magnetic Sensing inside a Diamond Anvil Cell via Nitrogen-Vacancy Center Spins*. *Chin Phys Lett* 2019;**36**(8):086201.

30  Yip KY, Ho KO, Yu KY *et al.* Measuring magnetic field texture in correlated electron systems under extreme conditions. *Science* 2019; **366**(6471):1355-59.

31  Bhattacharyya P, Chen W, Huang X *et al.* Imaging the Meissner effect in hydride superconductors using quantum sensors. *Nature* 2024;**627**(8002):73-9.

32  Wang M, Wang Y, Liu Z *et al.* Imaging magnetism evolution of magnetite to megabar pressure range with quantum sensors in diamond anvil cell. *Nat Commun* 2024;**15**(1):1-8.

33  Dai J-H, Shang Y-X, Yu Y-H *et al.* Optically Detected Magnetic Resonance of Diamond Nitrogen-Vacancy Centers under Megabar Pressures. *Chin Phys Lett* 2022;**39**(11):117601.

34  Wang M, Wen H-H, Wu T *et al.* Normal and Superconducting Properties of La$_3$Ni$_2$O$_7$. *Chin Phys Lett* 2024;**41**(7):077402.

35  Galven C, Mounier D, Bouchevreau B *et al.* Phase Transitions in the Ruddlesden–Popper Phase Li$_2$CaTa$_2$O$_7$: X-ray and Neutron Powder Thermodiffraction, TEM, Raman, and SHG Experiments. *Inorg Chem* 2016;**55**(5):2309-23.

36  Dias A, Viegas JI and Moreira RL Synthesis and µ-Raman scattering of Ruddlesden-Popper ceramics Sr$_3$Ti$_2$O$_7$, SrLa$_2$Al$_2$O$_7$ and Sr$_2$LaAlTiO$_7$. *J Alloys and Compd* 2017;**725**:77-83.

37  Adler P, Goncharov AF, Syassen K *et al.* Optical reflectivity and Raman spectra of Sr$_2$FeO$_4$ under pressure. *Phys Rev B* 1994;**50**(16):11396-402.

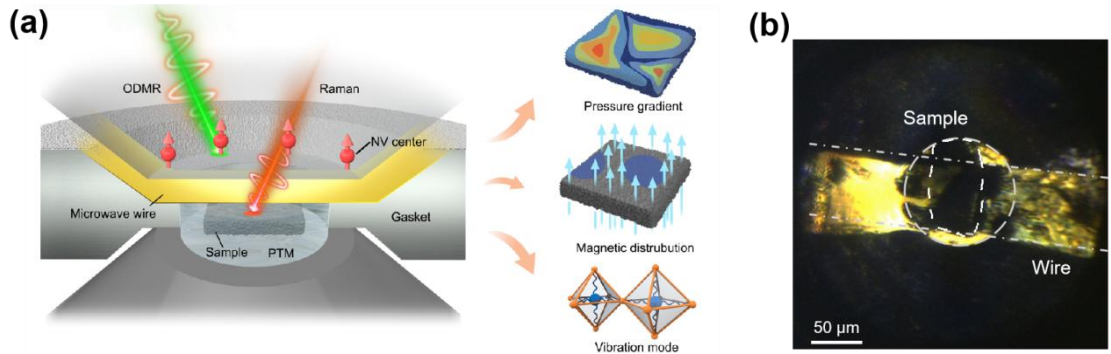

**Fig. 1 Correlated characterization of La$_3$Ni$_2$O$_{7-\delta}$ in diamond anvil cells (DACs). (a)** The La$_3$Ni$_2$O$_{7-\delta}$ and La$_2$PrNi$_2$O$_7$ samples are wrapped in the pressure transmitting medium (silicone oil or KBr) inside the DAC sample chamber. Correlated measurements of NV-based quantum sensing and Raman spectra are performed to reveal the pressure gradient, magnetic distribution and vibration mode of the samples. This provides a direct method to study the Meissner effect of superconductors under pressure. For in-situ magnetic field and pressure measurement, diamond anvils ([111]-crystal cut) with a layer of shallow NV centers are used, and a gold antenna is placed between the anvil culet and the pressure transmitting medium to transmit the microwave signal. (Deleted Fig. 1b which is about perfect conductor)**(b)** Bright field image of the diamond culet after loading sample A with 20 GPa. The sample is outlined by a white dashed line.

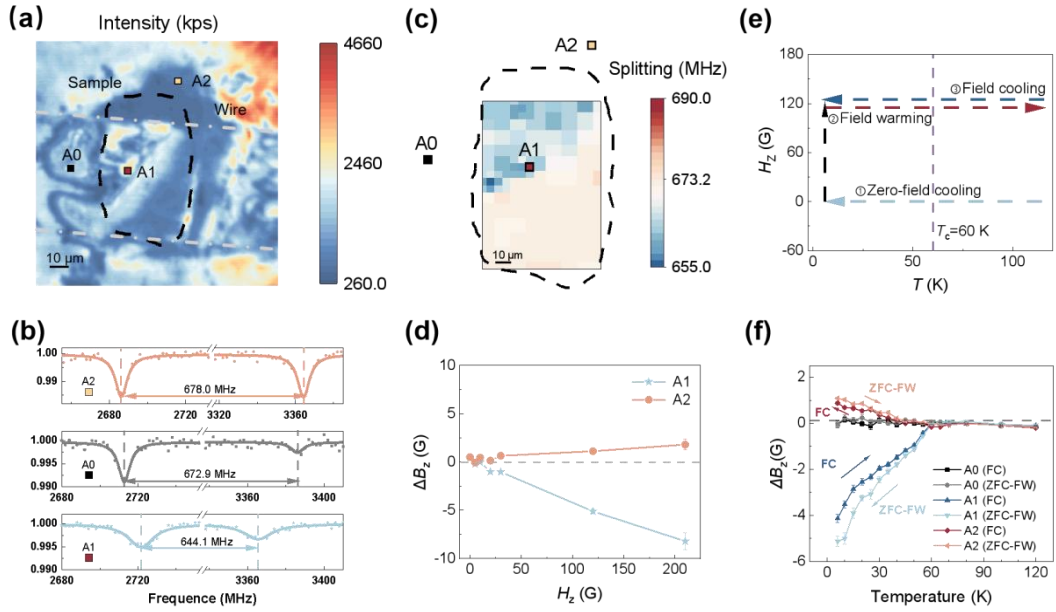

**Fig. 2 Local diamagnetism in $La_2PrNi_2O_7$ at 20 GPa. (a)** Fluorescence image of sample A ($La_2PrNi_2O_7$ in silicon oil). The edge of the sample is outlined with a black dashed line. **(b-c)** Typical ODMR spectra of the NV centers **(b)** and magnetic field mapping **(c)** under an external magnetic field of $H_Z \sim 120$ G after zero-field cooling of the sample to 7 K. Three points (A0, A1, A2) are selected based on their position with respect to the $La_2PrNi_2O_7$ sample. Point A1 is directly on the sample, A2 is at the edge of the sample and A0 is far away from the sample (serves as a reference point). The blue area in **(c)** shows the local diamagnetism (these points have a smaller ODMR splitting than that of the reference point). **(d)** Local magnetic field at the NV positions as a function of the applied external magnetic field (after zero-field cooling). For the NV centers at point A1, their local magnetic field is about 5% smaller than the applied external magnetic field. In sharp contrast, the NV centers at point A2 feel an enhanced local magnetic field. **(e)** Measurement protocol of field warming after zero-field cooling (ZFC-FW) and field cooling (FC). **(f)** ODMR splittings of three selected points under 120 G ZFC-FW and FC measurements. The combination of the ZFC-FW and FC curves provide clear evidence of the Meissner effect.

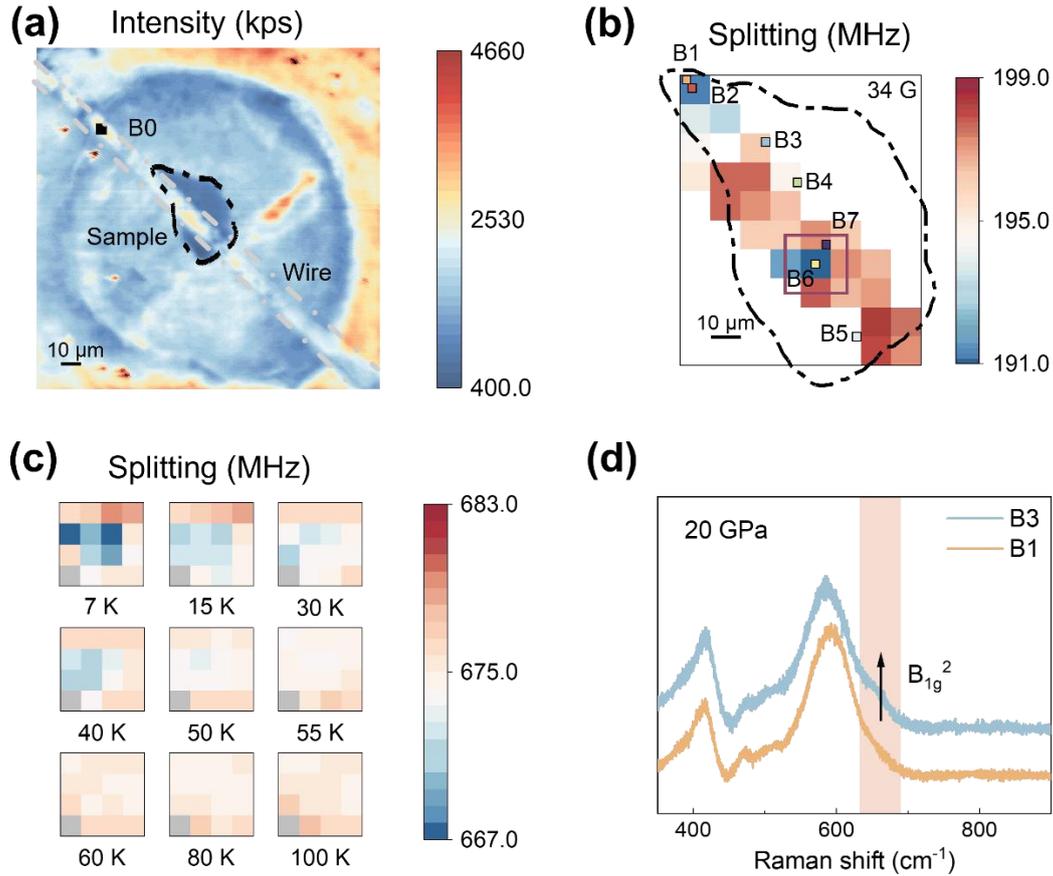

**Fig. 3 Correlated magnetic and Raman measurements of La$_2$PrNi$_2$O$_7$ under pressure. (a)** Fluorescence image of sample B (La$_2$PrNi$_2$O$_7$ in KBr). The edge of the sample is outlined with a black dashed line. **(b)** Magnetic field mapping under an external magnetic field of $H_Z$ = 34 G after zero-field cooling of the sample to 7 K. The blue region exhibits a noticeable diamagnetism. In-situ Raman measurements are performed on the superconducting regions (B1, B2, B4) and the non-superconducting regions (B3, B5) under different pressures. **(c)** Magnetic field mapping of one of the superconducting regions (position of purple square marked in b) during the FW process, the external magnetic field is around 120 G. Above the critical temperature $T_c$, the local diamagnetism disappears and all NV centers feel the uniform external magnetic field. The region on the gray square shows a feeble contrast to fit the ODMR splitting. **(d)** The Raman spectrum of B1 and B3 at 20 GPa. As shown in the upper part of the figure, a small satellite peak is observed in the non-superconducting region (B3) at 20 GPa, while it is suppressed in the superconducting region (B1). This peak is also presented at atmospheric pressure before compression (shown in Fig. S6a).

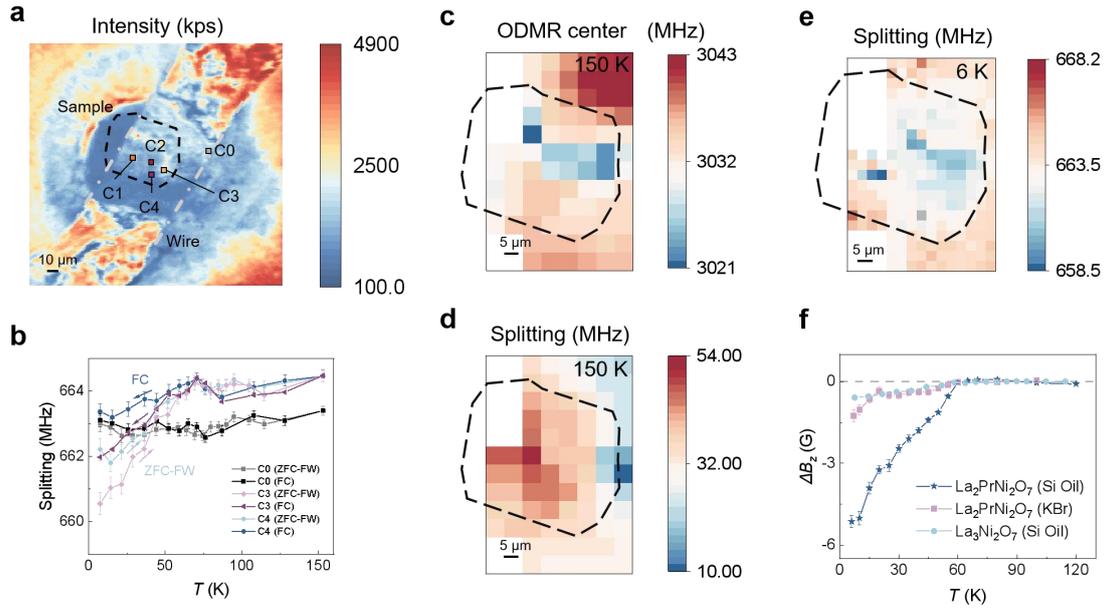

**Fig. 4 Magnetic and stress measurement of the $La_3Ni_2O_{7-\delta}$ sample at 20 GPa. (a)** Fluorescence image of sample C ($La_3Ni_2O_{7-\delta}$ in silicon oil). The edge of the sample is outlined by the black dashed line. **(b)** ODMR splitting of the NV centers during the ZFC-FW and FC processes, with an external magnetic field of around 120 G. The NV positions are marked in **(a)**. Both the diamagnetism of the $La_3Ni_2O_{7-\delta}$ sample and the local stress contribute to the ODMR splitting. **(c-d)** Stress distribution revealed by the zero-field ODMR spectra at 150 K ($> T_C$). The center frequency of the ODMR spectra **(c)** reveals the compressive stress, while the splitting of the ODMR spectra **(d)** is proportional to the differential stress. **(e)** Magnetic image under an external magnetic field of around 120 G after zero-field cooling. Note that the contribution of the stress distribution has been subtracted by the results shown in **(c-d)**. The region on the gray square is an odd point and is therefore not included in the discussion. **(f)** Comparison of the diamagnetism effect of the three samples during the ZFW-FW measurement. The external magnetic field is around 120 G. For each point, the ODMR splitting at high temperatures ($> T_C$) is used as the reference to calculate $\Delta B_Z$.

15